\documentclass[twocolumn,superscriptaddress]{revtex4-1}
\usepackage{amsmath}%
\usepackage{bm}%
\usepackage{graphicx}
\usepackage{color}
\begin{document}

\title{Evolution of the partially frustrated magnetic order in CePd$_{1-x}$Ni$_x$Al}

\author{Zita Huesges}
\affiliation{Max-Planck-Institut f\"ur Chemische Physik fester Stoffe, 01187 Dresden, Germany}
\affiliation{Helmholtz-Zentrum Berlin f\"ur Materialien und Energie GmbH, 14109 Berlin, Germany}
\author{Stefan Lucas}
\affiliation{Max-Planck-Institut f\"ur Chemische Physik fester Stoffe, 01187 Dresden, Germany}
\author{Sarah Wunderlich}
\affiliation{Max-Planck-Institut f\"ur Chemische Physik fester Stoffe, 01187 Dresden, Germany}
\author{Fabiano Yokaichiya}
\affiliation{Helmholtz-Zentrum Berlin f\"ur Materialien und Energie GmbH, 14109 Berlin, Germany} 
\author{Karel Proke\v{s}}
\affiliation{Helmholtz-Zentrum Berlin f\"ur Materialien und Energie GmbH, 14109 Berlin, Germany}
\author{Karin Schmalzl}
\affiliation{J\"ulich Centre for Neutron Science JCNS, Forschungszentrum J\"ulich GmbH, Outstation at ILL, 38042 Grenoble, France}
\author{Marie-H\'{e}l\`{e}ne Lem\'{e}e-Cailleau}
\affiliation{Institut Laue-Langevin, 38042 Grenoble, France}
\author{Bj\o rn Pedersen}
\affiliation{Heinz Maier-Leibnitz Zentrum (MLZ), TU M\"{u}nchen, 85748 Garching, Germany}
\author{Veronika Fritsch}
\affiliation{Experimentalphysik VI, Zentrum f\"{u}r Elektronische Korrelationen und Magnetismus, Institut f\"{u}r Physik, Universit\"{a}t Augsburg, 86135 Augsburg, Germany}
\author{Hilbert v. L\"ohneysen}
\affiliation{Karlsruher Institut f\"{u}r Technologie, Physikalisches Institut und Institut f\"{u}r Festk\"{o}rperphysik, 76131 Karlsruhe, Germany}
\author{Oliver Stockert}
\affiliation{Max-Planck-Institut f\"ur Chemische Physik fester Stoffe, 01187 Dresden, Germany}

\date{\today}

\begin{abstract}
We report on a single-crystal neutron diffraction study of the evolution of the antiferromagnetic order in the heavy-fermion compound CePd$_{1-x}$Ni$_x$Al which exhibits partial geometric frustration due to its distorted Kagom\'{e} structure. The magnetic structure is found to be unchanged with a propagation vector $Q_\mathrm{AF} \approx (0.5~0~0.35)$ for all Ni concentrations $x$ up to $x_c \approx 0.14$. Upon approaching the quantum critical concentration $x_c$, the ordered moment vanishes linearly with N\'eel temperature $T_{\rm N}$, in good agreement with CePdAl under hydrostatic pressure. For all Ni concentrations, substantial short-range magnetic correlations are observed above $T_{\rm N}$ as a result of frustration.
 
\end{abstract}

\maketitle

Competing interactions in condensed matter systems attract both experimentalists and theorists due to the large variety of resulting phases and new physics involved. Depending on the relative strength of the interactions, different ground states can occur. In particular, metallic systems close to a magnetic instability at zero temperature, i.e., a quantum critical point in the case of a continuous transition, are interesting: Here the quantum-critical fluctuations of the order parameter determine the thermodynamic and transport properties also at finite temperatures resulting, e.g., in non-Fermi-liquid behavior \cite{Loehneysen2007,Si2001,Abrahams2014}. Heavy-fermion compounds are particularly well-suited model systems to study quantum criticality because the strength of the competing interactions can be tuned by, e.g., alloying, hydrostatic pressure, or magnetic field. Recently, magnetic frustration was suggested as a further parameter to tune quantum criticality \cite{Si2006,Vojta2008,Coleman2010}.

The heavy-fermion compound CePdAl combines both the vicinity to a quantum critical point in a metallic magnet and the appearance of geometric magnetic frustration. Although at first sight magnetic frustration and heavy-fermion magnetism seem to be contradictory, the magnetic $4f$ moments are neither fully localised nor itinerant and therefore allow for frustration in a metallic system. CePdAl crystallises in the hexagonal ZrNiAl structure with the cerium atoms located on a distorted Kagom\'{e} lattice in the hexagonal basal plane \cite{Xue1994}. It shows the typical properties of a heavy-fermion compound with an enhanced Sommerfeld coefficient of the specific heat being $\gamma \approx 270$\,mJ/molK$^2$ \cite{Schank1994}, and orders antiferromagnetically below the N\'eel temperature $T_{\rm N} = 2.7$\,K in an incommensurate magnetic structure with a propagation vector $Q_\mathrm{AF} \approx (0.5~0~0.35)$ as revealed by neutron diffraction \cite{Doenni1996,Keller2002,Prokes2006}. The cerium magnetic moments of $\approx 1.7\,\mu_{\rm B}$ are aligned along the $c$ direction \cite{Doenni1996,Keller2002,Prokes2006} which is the easy axis in this Ising-like system. Notably, the neutron diffraction data have shown that only $2/3$ of the cerium moments participate in the antiferromagnetic state, while $1/3$ of the moments remains disordered down to lowest temperatures---despite the fact that all cerium atoms occupy crystallographically equivalent sites. Measurements in a field $B \parallel c$ revealed a rich $B-T$ phase diagram, triggered by the interplay of the response of ordered and non-ordered Ce moments \cite{Prokes2006,Zhao2016,Mochidzuki2017,Lucas2017}, and thus give further evidence of the importance of partial frustration for the properties of CePdAl.

CePdAl can be tuned to a quantum critical point either by applying hydrostatic pressure \cite{Tang1996,Goto2002,Prokes2007} or by composition \cite{Fritsch2014,Sakai2016}. Substituting isoelectronic Ni for Pd leads to a decrease of the magnetic transition temperature and at a critical Ni content of $x_c \approx 0.14$ a quantum critical point, i.e., $T_{\rm N} = 0$, is approached with the appearance of non-Fermi-liquid behaviour \cite{Fritsch2014,Sakai2016}. In particular, the N\'eel temperature in CePd$_{1-x}$Ni$_x$Al varies linearly with Ni concentration, viz. $T_{\rm N} \propto |x-x_c|$. The crystal structure is stable through the substitution series, with only small changes of the lattice constants which are linear in $x$ (for CePdAl, $a =$ \mbox{7.21\,\AA} and $c =$ \mbox{4.23\,\AA}) \cite{Fritsch2013}. So far, nothing has been known about the evolution of the magnetic structure in CePd$_{1-x}$Ni$_x$Al and about possible changes to the frustration upon approaching quantum criticality. Neutron scattering is an indispensable method to probe magnetic ordering phenomena on a microscopic scale. Hence, in the following we report on a detailed neutron scattering study of the magnetic order of CePd$_{1-x}$Ni$_x$Al single crystals.

Neutron diffraction experiments were performed on three different Czochralski-grown CePd$_{1-x}$Ni$_x$Al single crystals with $x = 0, 0.05$ and $0.1$ ($m \approx 1.8$, $3$, and $7$\,g). The samples were measured with identical setup on the single-crystal diffractometer E4 at the BER-II neutron source of the Helmholtz-Zentrum Berlin \cite{Prokes2017} using a neutron wavelength $\lambda = 2.41$\,{\AA}. In addition, the pure CePdAl crystal was measured in detail on the diffractometers D10 at the high-flux reactor of the Institut Laue-Langevin in Grenoble ($\lambda = 2.36$\,{\AA}) and on RESI at the neutron source of the Maier-Leibnitz-Zentrum in Garching \cite{Pedersen2015} ($\lambda = 1.03$\,{\AA}). Furthermore, elastic neutron scattering was carried out on a nominal CePd$_{0.86}$Ni$_{0.14}$Al single crystal ($m \approx 6.2$\,g) at the triple-axis spectrometer IN12 of the Institut Laue-Langevin. Here a neutron wavelength of $\lambda= 5.46$\,{\AA} was used. All crystals were mounted with the $[110]$ axis vertical resulting in a $(h~0~l)$ horizontal scattering plane. Data were taken in the temperature range up to $10$\,K using $^3$He cryostats (base temperature $\approx 0.25$\,K, at E4 and RESI), a $^4$He cryostat (down to $\approx 1.6$\,K, at D10), or a dilution cryostat (base temperature $\approx 0.06$\,K, at IN12). All data were normalised to a monitor placed in the incident neutron beam. Typical counting times per point were of the order of a minute, typical intensities around 5 counts per second in the magnetic peak (at E4). $Q$ scans were taken along $h$ and $l$, and at D10, where out-of-plane measurements were possible, also along $k$.
 
Energy-dispersive x-ray (EDX) measurements were performed on all samples and confirmed the nominal Ni concentration for $x = 0$, 0.05 and 0.1. However, for the sample with nominal $x = 0.14$, the EDX data suggest that the true Ni concentration is lower, $x \approx 0.135$, so that the sample is located slightly below the quantum critical point $x_\mathrm{c} = 0.14$ and has a finite transition temperature and ordered moment. Moreover, a concentration gradient of $\pm$0.01 exists in this sample, so that some parts are sub-critical, i.e., magnetically ordered, while others might remain paramagnetic.

\begin{figure}
\centering
\includegraphics[width=\linewidth]{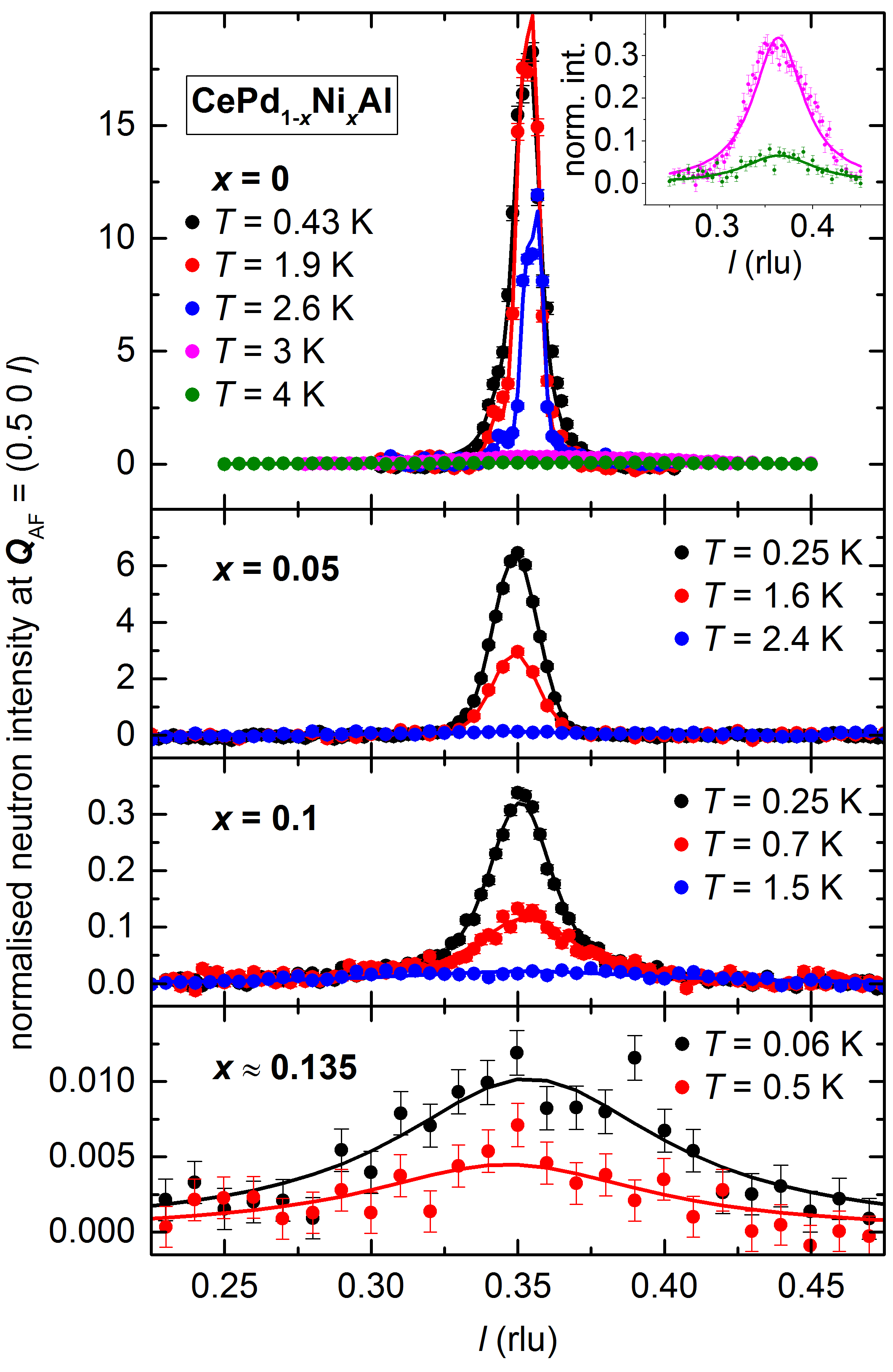}
\caption{$Q$ scans along $[001]$ across the position of the magnetic superstructure peak at $Q_\mathrm{AF} \approx (0.5~0~0.35)$ for different samples CePd$_{1-x}$Ni$_{x}$Al. Solid lines denote fits of the peaks (see main text). All data, measured at RESI, D10, E4 and IN12, are normalised to the nuclear reflections so that the intensities can be compared. The background has been subtracted for better visibility. A few selected temperatures are shown for each sample. In the inset, scans of CePdAl measured above the N\'{e}el temperature are displayed on an expanded vertical scale.}
\label{lscans}
\end{figure}

Fig.\,\ref{lscans} shows selected scans along [001] across a magnetic Bragg peak of CePdAl at $Q_\mathrm{AF} \approx (0.5~0~0.35)$ for different temperatures. Further magnetic reflections were also measured; they all are in good agreement with the ordering wave vector $(0.5~0~\tau)$, $\tau \approx 0.35$, which has been deduced from powder \cite{Doenni1996,Keller2002} and single-crystal \cite{Prokes2006} neutron diffraction data. We also observe that the incommensurate component $\tau$ is temperature dependent (cf. Fig.\,\ref{lscans}) between $T_{\rm N}$ and the lock-in temperature 1.9\,K, again in accordance with previous work \cite{Keller2002,Prokes2006}. 
 
For CePd$_{1-x}$Ni$_x$Al with $x = 0.05$ and $0.1$, we observe magnetic reflections with the same ordering wave vector as for the pure compound (Fig.\,\ref{lscans}). In the sample with Ni content $x \approx 0.135$ we find broad correlations at the same wave vector. Thus, Ni-substituted CePdAl behaves similar to CePdAl under pressure, where it was also found that the ordering wave vector remains unchanged \cite{Prokes2007}.

The magnetic peaks were fitted with Voigt functions: The Lorentzian component describes the correlation length or domain size (termed characteristic length because the distinction is not clear below $T_{\rm N}$), while the convolution with a Gaussian profile accounts for the experimental resolution. For pure CePdAl, all reflections including the nuclear reflections are split into three peaks due to a slight misalignment of crystallites in the sample. This is accounted for by our fit function, which is a sum of three peaks. For the Ni-substituted samples, there are several crystallites whose reflections cannot be separated with confidence, so we describe the scans with a single Voigt peak and absorb the effect of mosaicity into the Gaussian resolution. Thus, the effective resolution is limited by the instrumental setup for pure CePdAl and, instead, by the sample quality for all samples with $x > 0$. The reduced $Q$ resolution hampers a precise determination of the incommensurate component $\tau$ and thus of the lock-in temperature (if existent) for the Ni-substituted samples. 
\begin{figure}
\centering
\includegraphics[width=\linewidth]{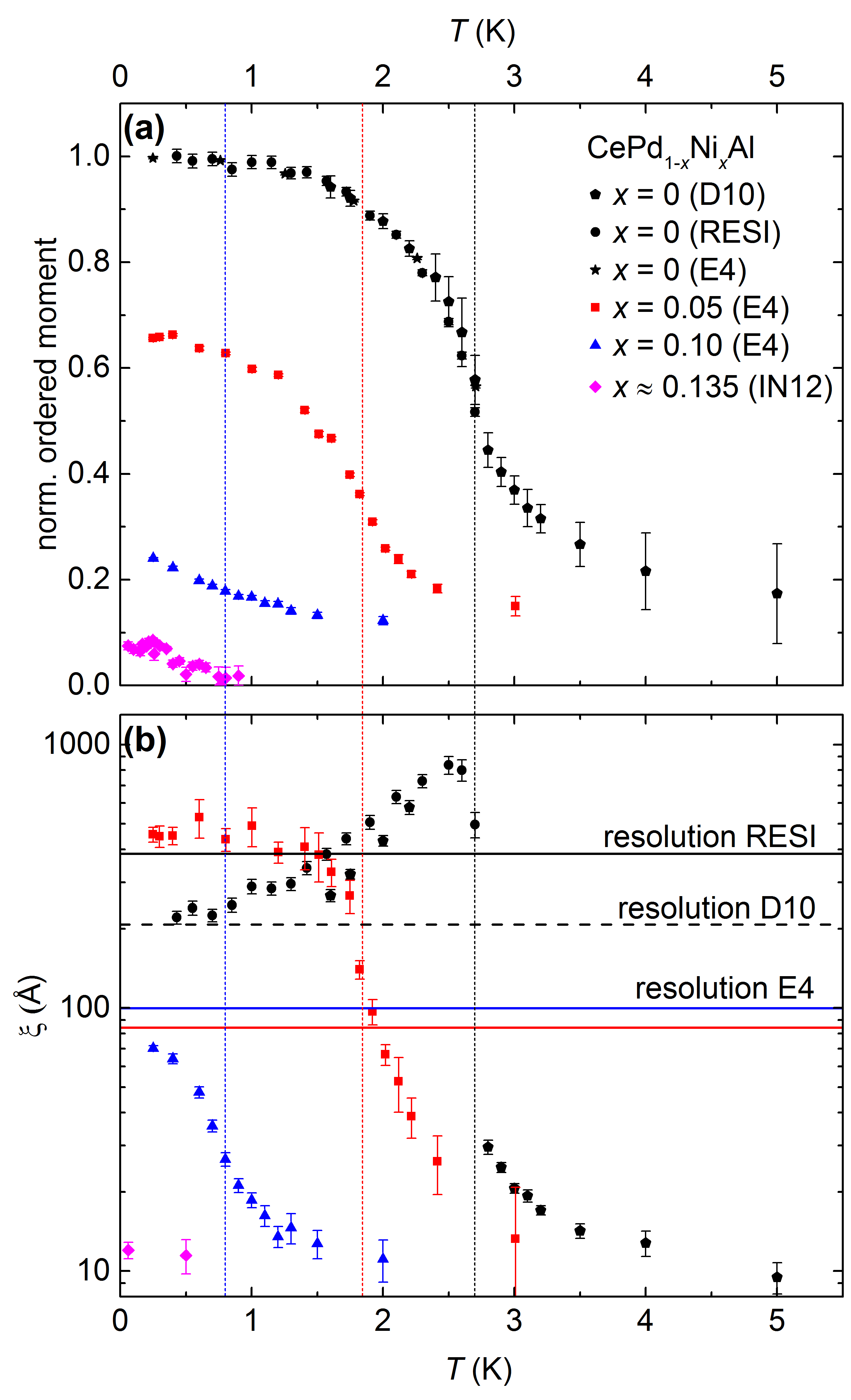}
\caption{Temperature dependence of (a) the ordered magnetic moment $m$ and (b) the characteristic length $\xi$, on a logarithmic scale, in CePd$_{1-x}$Ni$_x$Al with different Ni concentrations. The ordered moments are normalised to the extrapolated value of pure CePdAl for $T \to 0$. The dashed vertical lines indicate the N\'{e}el temperatures as determined from the neutron data (see main text for details). The data shown here are deduced from fits to scans along $[001]$ across the $(0.5~0~0.35)$ reflection. An exception is the ordered moment of the sample with $x \approx 0.135$, where most points correspond to counting at the centre of the peak, since only few $Q$ scans were measured. The horizontal lines in panel (b) show the instrument resolution (D10, RESI), or the effective resolution (E4) given by the sample mosaicity.}
\label{Tdep}
\end{figure}

Figs.\,\ref{Tdep} (a) and (b) display the ordered magnetic moment $m$ and the characteristic length, respectively, of all CePd$_{1-x}$Ni$_x$Al samples as a function of temperature, as deduced from Voigt fits to $Q$ scans across the $(0.5~0~0.35)$ reflection. Since the magnetic structure remains unchanged, the normalised ordered moment can be calculated as the square root of the integrated magnetic intensity. The ordered moments per Ce atom are normalised to the value in pure CePdAl for $T \to 0$, which is \mbox{$\approx$ 1.7\,$\mu_\mathrm{B}$} \cite{Doenni1996,Keller2002,Prokes2006}. The scaling between the samples is based on the averaged intensity of several nuclear reflections for each sample. For the sample with $x \approx 0.135$, measured at IN12, fewer nuclear reflections were	accessible, so that the intensity scaling is less precise. We define the characteristic length as $\xi = 1/\kappa$, $\kappa$ being the half width at half maximum of the Lorentzian magnetic peak in units of \AA$^{-1}$. For comparison, the (effective) resolution (the inverse of the Gaussian half width) is also given in Fig.\,\ref{Tdep}.

The dashed vertical lines in Fig.\,\ref{Tdep} show the N\'{e}el temperatures as deduced from the neutron data. For $x = 0$ and $0.05$, clear signatures of the N\'{e}el temperature are observed, for both the ordered moment and for the characteristic length, and the agreement with N\'{e}el temperatures found in heat-capacity measurements \cite{Sakai2016} is very good. For these samples, $T_{\rm N}$ can be unequivocally taken as the inflexion point of $m(T)$. However, for $x = 0.1$ and $0.135$ a gradual onset of magnetic order is observed, and the order does not truly become long-ranged even at lowest measuring temperatures. The N\'{e}el temperature found for the $x = 0.1$ sample, $T_{\rm N} =$ 0.8\,K (visible as a weak kink of $m(T)$), agrees fairly well with the heat-capacity data, while for the $x \approx 0.135$ sample it is not possible to extract a transition temperature from the neutron data. Similarly, in the thermodynamic measurements, the transition into the ordered state is sharp and $\lambda$-like for small $x$, but becomes very broad for $x \approx 0.1$ and larger \cite{Fritsch2014,Sakai2016}. CePdAl under hydrostatic pressure also shows broader transitions in the heat-capacity data at higher pressures \cite{Goto2002}, however, the peaks are sharper than for the Ni-substituted samples when comparing samples with similar N\'{e}el temperatures. Thus, local disorder might be partially responsible for the transition broadening observed in our data.

The shape of the $m(T)$ curves of CePd$_{1-x}$Ni$_x$Al, as extracted from neutron scattering, is atypical compared to other heavy-fermion systems approaching quantum criticality such as CeCu$_{6-x}$Au$_x$ \cite{Loehneysen1998} or CePd$_2$Si$_2$ under pressure \cite{Kernavanois2005}: The staggered magnetisation at $T_{\rm N}$ and above is unusually high, e.g. $m(T_{\rm N})\approx 0.5\cdot m(T \to 0)$ and $m(1.5~T_{\rm N}) \approx 0.25\cdot m(T \to 0)$ for CePdAl and CePd$_{0.95}$Ni$_{0.05}$Al. This is not simply due to transition broadening as evidenced by the following: (1) It is observed also in CePdAl and CePd$_{0.95}$Ni$_{0.05}$Al, which show sharp transitions into the ordered state in heat-capacity data; (2) the correlations persist to high temperatures, i.e., they are measurable even at temperatures twice as high as the N\'{e}el temperature; (3) at $T \approx 1.5~T_{\rm N}$, we find a similar characteristic length of around 12-14\,{\AA} for all measured samples. We relate the observed increase of (quantum) critical fluctuations to the presence of frustration in CePd$_{1-x}$Ni$_x$Al. 

Prior neutron measurements of CePdAl did not observe short-range correlations, probably because the signal-to-background ratio is lower for both powder measurements \cite{Doenni1996,Keller2002} and single-crystal measurements with pressure cells \cite{Prokes2007}. However, short-range correlations above $T_{\rm N}$ were reported based on NMR measurements of pure CePdAl by Oyamada \emph{et al.} \cite{Oyamada2008}. The authors estimated the correlation length at 3\,K to be $2.3\cdot a =$ 16.5\,{\AA}, which was measured in powder samples so that the reference to the $a$ axis is only an assumption. Our single-crystal neutron measurement yields 15.8\,{\AA} $= 2.2\cdot a$ along $h$ (real-space direction $[1\bar{1}0]$), 12.6\,{\AA} $= 1.7\cdot a$ along $k$ ($[\bar{1}10]$) and 21.4\,{\AA} $= 5.1\cdot c$ along $l$ ($[001]$) at 3\,K, in very good overall agreement with the NMR data. In thermodynamic properties of pure CePdAl, an accumulation of entropy above $T_{\rm N}$ has recently been seen \cite{Lucas2017}, which has been related to the short-range correlations. For CePd$_{1-x}$Ni$_x$Al with 0.05 $\le x \le$ 0.16, a contribution to the magnetic Gr\"{u}neisen parameter near 2.5\,T was ascribed to correlations related to frustration \cite{Sakai2016}. 

A peculiarity is observed in the ordered phase of pure CePdAl: The peak width increases towards lowest temperatures, i.e., the characteristic length decreases significantly (Fig.\,\ref{Tdep} (b)), while sharp, resolution-limited peaks are observed in the vicinity of $T_{\rm N}$. At the same time, the temperature dependence of the peak intensity behaves normally (Fig.\,\ref{Tdep} (a)). D\"{o}nni \emph{et al.} \cite{Doenni1996} and Proke\u{s} \emph{et al.} \cite{Prokes2007} observed that magnetic peaks are broader than nuclear peaks in CePdAl, but they did not report a temperature dependence of this observation. The fact that the peaks are sharper at intermediate temperatures clearly rules out disorder as an origin of the broadening. Instead, this unusual feature might be related to the presence of competing magnetic interactions, with the result that correlations of a competing interaction limit the range of the order with $Q_\mathrm{AF} \approx (0.5~0~0.35)$ (cf. also ref. \cite{Fritsch2017}). 

For the Ni-substituted samples, an analogous behaviour could not be observed. In the case of $x = 0.05$, the observation might be hindered by the experimental resolution. For the $x = 0.1$ and $x \approx 0.135$ samples, the magnetic peaks are broader than the resolution at all temperatures. Whether this broadening in the ordered state is related to disorder or frustration cannot be decided on the basis of our data.

\begin{figure}
\centering
\includegraphics[width=0.9\linewidth]{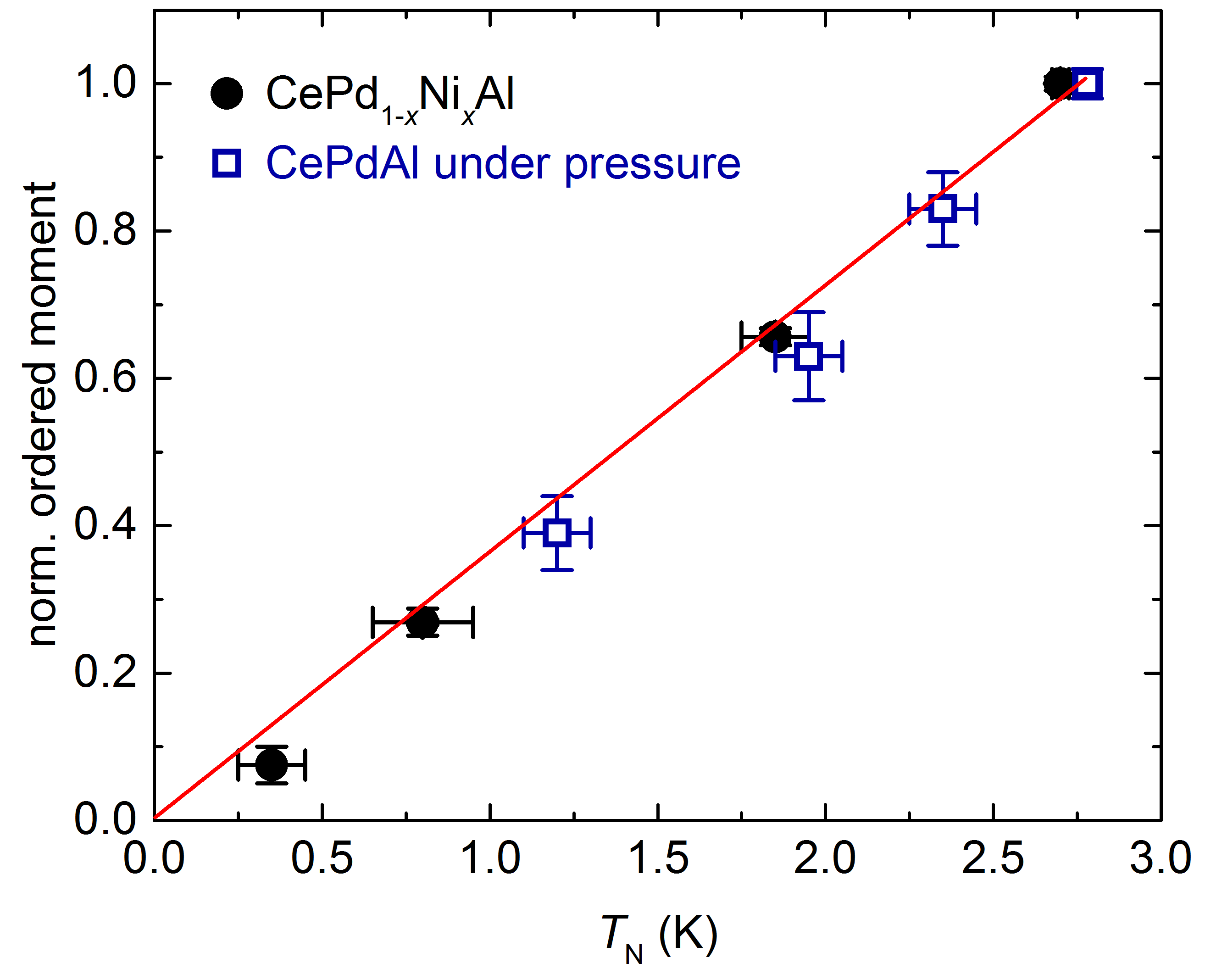}
\caption{Ordered magnetic moment $m$ (extrapolated to $T = 0$) versus N\'eel temperature $T_{\rm N}$ of CePd$_{1-x}$Ni$_x$Al ($x =$ 0, 0.05, 0.1 and 0.135), as well as of CePdAl under hydrostatic pressure ($p =$ 0, 0.4, 0.55 and 0.65\,GPa) \cite{Prokes2007,comment}. The ordered moments are normalised to the value of $x = 0$ and $p = 0$, respectively. For the sample with $x \approx 0.135$, we determine the N\'{e}el temperature from a heat-capacity measurement of a small part of the crystal.}
\label{moment_vs_TN}
\end{figure}

For an accurate comparison of the ordered moment at $T \to 0$ of samples with different Ni concentration, a set of several magnetic peaks has been measured at base temperature for each sample. The averaged integrated magnetic intensity has then been normalised to the averaged intensity of several nuclear peaks, as described above. Again the data point of the $x \approx 0.135$ sample is less accurate because fewer reflections were accessible and a different instrument was used. In Fig.\,\ref{moment_vs_TN}, we show the ordered moments, extrapolated to $T = 0$, of all CePd$_{1-x}$Ni$_x$Al samples studied in this work, plotted as a function of $T_{\rm N}$. They are compared with data of pure CePdAl under hydrostatic pressure \cite{Prokes2007,comment}. The data sets are in good agreement with each other, confirming that the suppression of magnetic order in CePd$_{1-x}$Ni$_x$Al is a volume effect and not related to changes of the electronic structure induced by Ni substitution. A similar equivalence of composition and pressure tuning has been observed for CeCu$_{6-x}$Au$_x$ \cite{Hamann2013}. Thus, we believe that the only significant difference between CePd$_{1-x}$Ni$_x$Al and pure CePdAl under pressure is that the transition into the ordered state is sharper for the latter case.

Fig.\,\ref{moment_vs_TN} suggests a linear relationship of moment and N\'{e}el temperature in CePdAl and CePd$_{1-x}$Ni$_x$Al. Since $T_{\rm N}$ is proportional to the Ni concentration $x$ \cite{Fritsch2014,Sakai2016}, plotting the ordered moment as a function of $x$ rather than $T_{\rm N}$ also results in a straight line. A plot of the ordered moment versus pressure, as shown in Fig.\,2 of reference \cite{Prokes2007}, does not appear linear, most likely due to the limited accuracy of the pressure measurement. 

\begin{figure}
\centering
\includegraphics[width=0.9\linewidth]{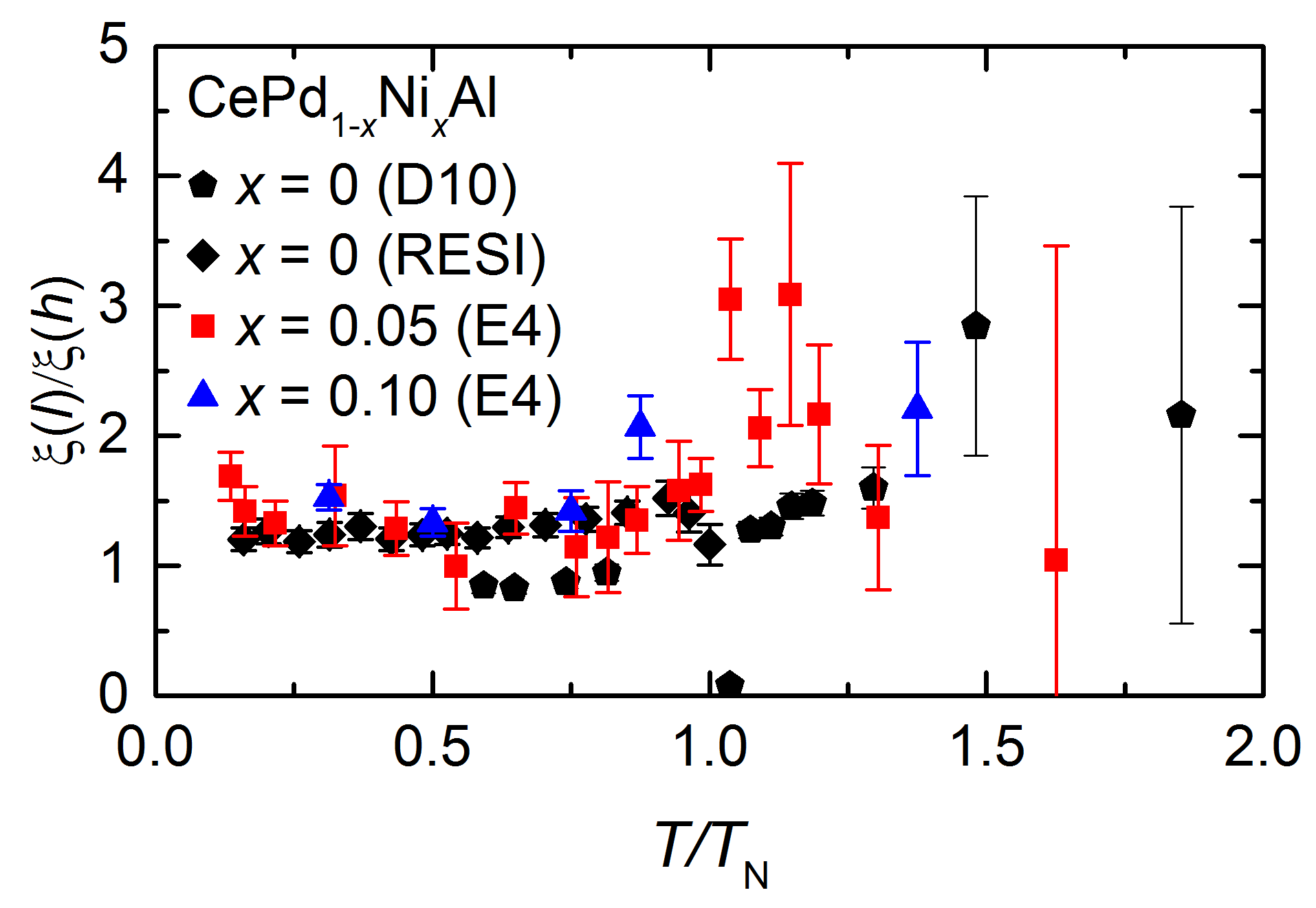}
\caption{Ratio of the characteristic length $\xi(l)$, as plotted in Fig.\,\ref{moment_vs_TN}, to the same quantity as extracted from $h$ scans, $\xi(h)$. The ratio is plotted for different CePd$_{1-x}$Ni$_x$Al samples as a function of reduced temperature $T/T_\mathrm{N}$. For $x \approx 0.135$, $h$ scans were not measured.}
\label{ratiolh}
\end{figure}

The linear dependence of N\'{e}el temperature and tuning parameter has been interpreted in terms of two-dimensional quantum critical behaviour within the Hertz-Millis-Moriya model, in conjunction with heat-capacity and magnetocaloric measurements \cite{Fritsch2014,Sakai2016}. However, in our neutron scattering experiments we find that the characteristic length is not strongly anisotropic, as evidenced by the ratio of the characteristic length of $l$ scans and $h$ scans in Fig.\,\ref{ratiolh}. A possible explanation for this apparent discrepancy might be that two-dimensional fluctuations arise at the quantum-critical endpoint of a  line of three-dimensional ordering temperatures, similar to the situation in quantum-critical CeCu$_{6-x}$Au$_x$ \cite{Stockert1998}. Alternatively, the observed critical exponents might have to be explained on the basis of new universality classes outside of the Hertz-Millis-Moriya model, possibly due to the influence of frustration.

In summary, using single-crystal neutron diffraction we were able to successfully study the evolution of magnetic order in the concentration series CePd$_{1-x}$Ni$_x$Al. The investigation of three different samples under equal conditions with the same diffractometer allows a reliable estimation of the evolution of the ordered magnetic moment, which scales linearly with the N\'{e}el temperature. A further sample, close to the critical concentration, was studied at a different instrument. All samples show the same ordering wave vector that relates to partially frustrated magnetic order. Frustration is further evident from strong correlations above $T_\mathrm{N}$ for the series CePd$_{1-x}$Ni$_x$Al. An unusual broadening of the magnetic reflections towards low temperatures is observed in pure CePdAl, possibly due to competing interactions. Our data provide microscopic evidence that frustration is present in the series CePd$_{1-x}$Ni$_x$Al right up to the quantum critical point. Further work is necessary to establish how it affects the quantum critical fluctuations.

We would like to acknowledge funding by DFG Research Unit 960 "Quantum Phase Transitions" and by the Helmholtz Association via VI-521. We thank U. Burkhardt for his support with the EDX measurements and are grateful for obtaining beam time at Helmholtz-Zentrum Berlin, Institut Laue-Langevin and Maier-Leibnitz Zentrum.
\bibliography{CePdAl_diffraction}

\end{document}